\documentstyle[aps]{revtex}
\input{epsf}
\begin{document}
\draft
\begin{title}
 {Two-Proton Correlations from 14.6A GeV/c Si+Pb and 11.5A GeV/c Au+Au
Central Collisions}
\end{title}
\author{J.~Barrette$^4$, R.~Bellwied$^{8}$, S.~Bennett$^{8}$, R.~Bersch$^6$, 
P.~Braun-Munzinger$^2$, W.~C.~Chang$^6$, W.~E.~Cleland$^5$, M.~Clemen$^5$, 
J.~Cole$^3$, 
T.~M.~Cormier$^{8}$, Y.~Dai$^4$, G.~David$^1$, J.~Dee$^6$, O.~Dietzsch$^7$,
M.~Drigert$^3$, K.~Filimonov$^4$, S.~C.~Johnson$^6$, S.~Gilbert$^4$, 
J.~R.~Hall$^8$, T.~K.~Hemmick$^6$,
N.~Herrmann$^2$, B.~Hong$^2$, C.~L.~Jiang$^6$, Y.~Kwon$^6$,
R.~Lacasse$^4$, Q.~Li$^{8}$, T.~W.~Ludlam$^1$,
S.~K.~Mark$^4$, R.~Matheus$^{8}$, S.~McCorkle$^1$, 
J.T.~Murgatroyd$^8$, D.~Mi\'skowiec$^2$,  E.~O'Brien$^1$,
S.~Y.~Panitkin$^6$, V.~S.~Pantuev$^6$, P. Paul$^6$, T.~Piazza$^6$, M.~Pollack$^6$, 
C.~Pruneau$^8$, M.~N.~Rao$^6$, E.~Reber$^3$, M.~Rosati$^4$, J.~Sheen{$^8$},
N.~C.~daSilva$^7$, S.~Sedykh$^6$, U.~Sonnadara$^5$,
J.~Stachel$^9$, H.~Takai$^1$, E.~M.~Takagui$^7$, S.~Voloshin$^{5,9}$,
T. ~Vongpaseuth$^6$, G.~Wang$^4$, J.~P.~Wessels$^9$, 
C.~L.~Woody$^1$, N.~Xu$^6$,
Y.~Zhang$^6$, Z.~Zhang$^5$, C.~Zou$^6$\\ (E814/E877 Collaboration)}
\address{$^1$ Brookhaven National Laboratory, Upton, NY 11973\\
$^2$ Gesellschaft f\"ur Schwerionenforschung, Darmstadt, Germany\\
$^3$ Idaho National Engineering Laboratory, Idaho Falls, ID 83402\\
$^4$ McGill University, Montreal, Canada\\
$^5$ University of Pittsburgh, Pittsburgh, PA 15260\\
$^6$ SUNY, Stony Brook, NY 11794\\
$^7$ University of S\~ao Paulo, Brazil\\
$^8$ Wayne State University, Detroit, MI 48202\\
$^9$ Physikalisches Institut der Universitaet Heidelberg, Heidelberg, Germany}

\date{\today}

\maketitle
\begin{abstract}
{Two-proton correlation functions have been measured in Si+Pb
collisions at 14.6
AGeV/c and Au+Au collisions at 11.5 AGeV/c  by the E814/E877 collaboration.
Data are compared with predictions of the transport model RQMD
and the source size is inferred from this
comparison. Our analysis shows that, for both reactions,
the characteristic size of the system at freeze-out exceeds the size of
the projectile, suggesting that the fireball created in
the collision has expanded.
For Au+Au reactions, the observed centrality dependence of the
two-proton correlation function implies that more central collisions
lead to a larger source sizes.
}
\end{abstract}

\section{Introduction}
Measurements of two-particle relative momentum correlations
\cite{Bauer92,Boal90} are widely used in relativistic heavy-ion
physics as a tool for extracting information about the spatial and
temporal extent of the system at freeze-out. 
The complexity of heavy-ion reactions
demands the utilization of different particles as probes of the reaction
zone in order to obtain a reliable picture of the collision. 
Most of the published experimental data are measurements of
two-meson ($\pi$ or K)
\cite{Abbot92,Akiba93,Barrette94,Alber,na44_kaon} correlation 
functions while the information concerning baryon freeze-out configurations 
is far more sparse. Several methods have been suggested to
address this problem: two-proton interferometry
\cite{Koonin77,Lednicky82}, analysis of the deuteron to proton yield
ratios \cite{Mekijan77} and deuteron-proton interferometry
\cite{Jennings86,Lednicky96}. 
In this paper we present measurements made by the E814/E877
collaboration at the BNL AGS of two-proton correlation functions in
Si+Pb collisions at 
14.6 AGeV/c and Au+Au collisions at 11.5 AGeV/c bombarding energy. The
results of the two-pion correlation analysis for these systems were
published in \cite{Barrette94,Barrette97}.
There, it was shown that pions at freze-out, i.e. when the system is
 dilute enough so that no further strong interactions take place,
 originate from a source which significantly exceeds in size the
 projectile. Although the pion source is, through the strong
 pion-nucleon interaction, closely coupled to the baryon dynamics, a
 direct measurement of the source size of baryons at freeze-out via
 two-proton correlations is clearly important.\\  
\indent Two-proton correlations are due to the
attractive strong and 
repulsive Coulomb final state interactions and are also influenced by
the effects of quantum statistics which requires an antisymmetrization of
the two-proton wave function. 
Coulomb repulsion, together with antisymmetrization, decreases the
probability of detection of pairs with relative momentum close to zero,
while the strong interaction increases this probability.
The interplay of these effects leads to a characteristic ``dip+bump''
shape of the correlation function. 
The height of the peak of the correlation function can be related to the
space-time parameters of the emitting source~\cite{Koonin77}. 
Thus an interpretation of the proton correlation data necessarily
involves calculations of the final state interactions and assumptions
about the properties of the emitting system. It has been shown
~\cite{Koonin77,Lednicky82} that, for 
simple static sources, the height, by which the peak deviates from unity, 
scales approximately inversely proportional to the source volume.  
However, in heavy-ion collisions, a
direct determination of the source parameters from the proton
correlation function is far from being straightforward.
The existence of collective effects
\cite{Barrette94,pbm95,na44_slope},
which leads to dynamical correlations between the momentum
and position of the emitted hadrons, may create a situation in which
the correlation function does not reflect the actual size of the
source \cite{Barrette94,Vossnack93}.  
Under these conditions, utilization of a static source model ~\cite{Awes95} to
generate the correlation and extract source parameters can be misleading. 
It seems necessary to utilize models, hydro-dynamical or transport,
which attempt to describe collective effects to interpret the
proton correlation functions. 
In order to deduce source parameters from the measured proton
correlation functions, a transport model was used in the present analysis.\\ 
\section{Experimental Setups}
\subsection{E814 Experimental Setup}
 Data used in the Si+Pb analysis were obtained with the E814 apparatus
during the 1991 run. Detailed description of the experimental set up
can be found in Ref.~\cite{Barrette94}. The apparatus consisted of several  
detector groups. A set of two high granularity calorimeters, Target
Calorimeter (TCAL) and Participant Calorimeter (PCAL), positioned
around the target area provided event 
by event measurements of transverse energy ($E_{T}$) production in
the interval of pseudo rapidity -2.0$<\eta<$4.7 . This
measurement was supplemented by the data from a uranium calorimeter,
located downstream of the forward spectrometer, which covered angles
close to zero degrees, thus providing the experiment with the capability of
almost 4$\pi$ measurement of $E_{T}$. Global event characterization was
also performed by a set of silicon multiplicity detectors each segmented into
512 pads and covering the pseudo rapidity range 0.8$<\eta<$3.9 . 
These detectors provided input for the trigger system. Note that
the trigger was based on global event characteristics such
as transverse energy production and charged particle multiplicity in a
wide interval of pseudo rapidity. A high-resolution forward magnetic
spectrometer was used for registration and identification of charged
particles. The spectrometer had a rectangular aperture, with the beam
passing through it, covering angles -115$<\theta_{x}<$14 mr in the
magnetic bend ( horizontal) plane, and -21$<\theta_{y}<$21 mr in
the perpendicular plane \cite{e814_90,e814_92}. The corresponding
coverage in proton transverse momentum ($p_t$) and rapidity (y) is
shown in Fig.~\ref{Fig:acceptance} . 
The 1991 configuration of the spectrometer consisted of a dipole magnet,
two drift chambers and two time of flight (TOF) hodoscopes.
The drift chambers were located at 7 (DC2) and 11.5 (DC3) meters
downstream of the target. The position resolution of the drift
chambers in the $x$ direction was approximately 250 $\mu$m and 300
$\mu$m, respectively.  
The hodoscopes were located at 12 and 31 meters downstream from the
target with a time of flight resolution of 200 and 350 ps
respectively. 
A group of detectors, located upstream of the target, consisting
of a set of plastic scintillators and silicon detectors,
provided information about beam particles and generated a start signal
for the TOF system.Singly charged particles were selected by cuts on
the amplitude of the signal in the TOF hodoscopes.
Simultaneous measurements of the particle rigidity and time of flight in
the spectrometer provided proton identification up to the momentum of
the beam.  In order to decrease multiple scattering and
improve momentum resolution most of the space between the detectors in
the forward spectrometer was filled with helium gas. The momentum
resolution, found from the Monte Carlo studies, was better than 4.1\%
and is limited by multiple scattering. The invariant
relative momentum (q$_{inv}$ defined below) resolution was estimated to be on the
order of 11 MeV/c for q$_{inv} \le 40$ MeV/c (the most important
region for the correlation measurement). 
\subsection{E877 Experimental Setup} 
The E877 apparatus was an upgrade of the E814 setup, designed to
study central Au+Au collisions at about 11 AGeV/c. For a more detailed
description see Ref.~\cite{Barrette97}. For the 1995 run a 
collimator was placed inside the opening 
of the PCAL. This determined the
geometrical acceptance of the forward spectrometer with angular
coverage -134$<\theta_{x}<$16 mr 
and -11$<\theta_{y}<$11 mr. Fig. 1 shows the acceptance range in
proton $p_t$ and rapidity. The vertical acceptance was
reduced compared to the E814 configuration 
in order to limit the multiplicity in the forward spectrometer, since
the average multiplicity in Au+Au collisions is approximately 
2.5 times higher than in Si+Pb collisions. In order to handle the
higher multiplicities, the forward spectrometer was upgraded with a
set of four multiwire proportional chambers (MWPC) positioned between
the two drift chambers. Also, a new TOF hodoscope, with time
resolution of about 80 ps, was installed about 12 meters downstream from the target,
behind DC3. Event centrality determination was provided by the
calorimeters. The forward spectrometer and the track reconstruction software
were capable of tracking up to 25 charged particles per event, though an
average event contains only about 7. The momentum resolution was
further improved by 
adding more helium bags, thus reducing the multiple scattering over a
larger part of the trajectories in the forward spectrometer. 
Monte Carlo studies of the momentum resolution 
showed that it was better than 3\% for protons and pions (see also
~\cite{Barrette97}). This 
resolution was found to be in 
good agreement with the momentum dependence of the measured width on
the proton mass peak. The corresponding resolution in relative
momentum $q_{inv}$ was estimated to be of the order of 7 MeV/c at low relative momenta.
\section{Data analysis}
To determine the two-proton correlation function $C_{2}$
experimentally we employ the definition
\begin{eqnarray} C_2(q_{inv}) =
\frac{N_{tr}(q_{inv})}{N_{bk}(q_{inv})} \hspace{0.3cm}, \end{eqnarray}
\noindent{where}                                            
\begin{eqnarray}                                                     
q_{inv} = \frac{1}{2} \sqrt{-(p_1^{\mu} - p_2^{\mu})^2}              
\end{eqnarray}                                                      
\noindent{is the half relative invariant momentum between the two identical    
particles with four-momenta $p_1^{\mu}$ and $p_2^{\mu}$. The
quantities  N$_{tr}$ and N$_{bk}$
are the ``true'' and ``background'' two-particle distributions
obtained by taking particles from the same and different events,     
respectively. Before constructing the correlation function, several cuts were
applied. First, protons were identified using the TOF and momentum
measurements. The contamination of other particles in the proton
samples is small ($<2\%$) and is neglected in the following. Beam momentum
protons were suppressed by a cut on rapidity $y \le 
3.1$. In order to overcome effects of two-track reconstruction
inefficiencies, a cut on the two-track  horizontal separation in the drift
chambers was applied.  The value of the cut has been chosen to be
approximately equal to twice the size of the drift cell of a chamber,
and was 12 mm  and  24 mm for DC2 and DC3, respectively. 
Monte Carlo studies showed that these cuts
effectively suppress distortions due to the close track reconstruction
inefficiencies. A requirement that the two protons did not share the same slat
of the TOF hodoscope has been imposed on pairs from both
distributions. The ``background'' distribution, N$_{bk}$, was obtained
using an event mixing method. A single proton was selected from one event
and then combined with other protons selected from different events to
generate the N$_{bk}$ distribution. The large statistics of the
N$_{bk}$ distribution ensures that statistical errors in the
correlation function are dominated by the statistics of the true
proton pairs. 
The background distribution was normalized such that in a range of
$q_{inv}$ from 100 MeV/c to 1000 MeV/c the number of mixed pairs was
equal to the number of true pairs.
\section{Results}
\subsection{Results for 14.6 A$\cdot$GeV/c Si+Pb collisions }
From one magnetic field polarity setting of the 1991 run, 230k events were
selected with a cut on transverse energy corresponding to the
upper 10\% of the geometric cross section for Si+Pb collisions. The
average number of reconstructed protons per event was about 1.4 .
After all cuts the N$_{tr}$ distribution contains 55k reconstructed
proton pairs. 
Figure~\ref{Fig:proton_distr} shows the rapidity and transverse
momentum distributions of the proton pairs for the
different ranges of $q_{inv}$. For each pair, all vectorial
quantities ($\vec p$ and $\vec p_t$) were defined as an average
of two vectors.  The rapidity of the pair
was defined as the average of rapidities of individual
particles. The most important region for correlations, i.e. that with
$q_{inv}<50$ MeV/c, is hatched in the Figure.  In order to give a
more quantitative characterization of the distributions,
Table~\ref{Tab:mean_val} shows a summary of the averaged values for
the distributions presented in Figure~\ref{Fig:proton_distr}. 
Note that pairs with small $q_{inv}$  are somewhat ``softer'', on
average, than the rest of the data set. 
 The corresponding correlation function C$_2$, as defined by Eq.(1), is
plotted in the upper panel of Fig.~\ref{Fig:results}.
The error bars reflect only statistical uncertainties. 
At relative momentum close to 20 MeV/c, a prominent proton-proton
resonance peak is evident.
 The position and the height of the peak is consistent with
the E802 preliminary two-proton correlation function\cite{stephans}
measured in 14.6 A$\cdot$GeV/c Si+Au collisions at
$\sigma_{trig}/\sigma_{geom} \approx 0.12$.
Note, however, that the E802 measurement was performed at mid-rapidity, while our
measurement is at more forward rapidities.                
\subsection{Results for 11.5A$\cdot$GeV/c Au+Au collisions }
 A total of 47 million events for Au+Au collisions were
analyzed. Tracking was performed for those events with transverse 
energy deposited in the PCAL which corresponds to approximately the
most central $22\%$ of the inelastic cross section. In order to allow
direct comparison with the correlation function measured in Si+Pb
reactions the correlation function for Au+Au dataset was first
determined with the same centrality cut, i.e. for the upper $10\%$ of
the inelastic cross section.  
It can be seen from Fig.~\ref{Fig:results} that under these conditions
the height of the peak 
of the two-proton correlation function for Au+Au collisions is
significantly smaller than in the Si+Pb case, indicating a larger
proton source size.   
Subsequently more cuts on centrality were applied
in the correlation analysis. Because of higher statistics in
this data set and better momentum resolution it was possible to apply
tighter cuts on centrality as well as finer $q_{inv}$ binning for 
the correlation function. For the
purpose of studying the centrality dependence of the 
correlation function we subdivided the available data into two subsets
with different centralities: the most central 4$\%$ (14 million
events, 5.3 million pairs) and 11$\%$-16$\%$ (11 million events, 3.1
million pairs) of the inelastic cross section. 
Figure~\ref{Fig:proton_distr} shows again the
rapidity and transverse momentum distributions of the proton pairs for
different ranges of $q_{inv}$. Table~\ref{Tab:mean_val} shows a 
summary of the average values for the distributions presented in
Fig.~\ref{Fig:proton_distr}. The definitions of the variables and 
cuts are identical to those used in Si+Pb data set analysis.
The average value of the rapidity distribution is
fairly similar for all intervals of $q_{inv}$. 
The average rapidity of proton pairs in the less central data set is
somewhat higher than the average rapidity of pairs in the most central $4\%$ sample. 
The average transverse momentum is the same in both
subsets. Figure~\ref{Fig:centrality_cuts_94} shows the corresponding correlation 
functions. One trend can be observed here: in the
peak region the correlation function for the most central data set
($4\%$) is  lower than the correlation function from the less central
($11\%$ to $16\%$ centrality) data set. As was discussed in~\cite{Koonin77}
such behaviour is consistent with the assumption of a larger
source size in the most central collision. It should be mentioned that
the change in the height of the correlation function is determined by
the interplay of the changes of the nuclear  
overlap volume and momentum distributions,
which may affect the height of the 
correlation function in opposite ways.   
Note that, in a framework of the model discussed
in~\cite{Koonin77,Lednicky82}, the higher average momentum of the
proton pairs in the $11-16\%$ centrality data should diminish the height
of the correlation function, whereas if the source size is smaller in the
less central collisions, this would tend to increase the height of the
peak of the correlation function.
\section{Model Comparison and Discussion}
In order to extract physical information from the measured correlation
functions, we carried out a 
study using the event generator RQMD(version 1.08)~\cite{rqmd1,rqmd2}. This
model describes 
classical propagation of the particles, together with quantum effects of
stochastic scattering and Pauli blocking. It includes color strings,
baryon and meson resonances, as well as finite formation time for
created particles. It has been successfully used to describe many
features of relativistic heavy-ion collisions
~\cite{Barrette94,Barrette97,E877qm96}. In this model a particle's
freeze-out position is defined 
as the point of the last strong interaction. The basic structure of
our approach to compare model and data is as follows: by taking the
freeze-out phase-space distribution generated by RQMD and propagating
the particles  through the experimental acceptance, accounting for the
resolution of the detectors, a subset of the phase-space points was 
obtained. Then the Koonin-Pratt method~\cite{Pratt1,Pratt94} was used to
construct the proton-proton correlation function. This method
provides a description of the final state interactions between two
protons and antisymmetrization of the their relative wave function. 
 The results of the calculations are shown as open symbols in both
panels of Fig.~\ref{Fig:results}. 
One can see that  agreement between the experiment and the model is
fairly good.  
Hence, it is of interest to determine the parameters of the proton
freeze-out source generated by the RQMD model. 
 The RQMD space-time distributions for 14.6 AGeV/c Si+Pb and 11.5 AGeV/c Au+Au
 collisions are shown in Fig.~\ref{Fig:rqmd}. All calculations were
 performed in  the nucleon-nucleon center of mass frame and a ``top 10\%''
 centrality cut was imposed on model events. The solid-lines and
 dashed-lines represent the 
 distributions in the experimental acceptance and close to mid-rapidity
 ($|y-y_{NN}| \leq 1$), respectively. Statistical parameters of these
 distributions are listed in the Table~\ref{Tab:rqmd}. The correlation
 functions combine the space-time information 
and it is a non-trival task to disentangle these related
contributions. However, microscopic models provide a valuable insight
into these phenomena and several interesting features can be inferred from
Fig.~\ref{Fig:rqmd}  and Table~\ref{Tab:rqmd}.  
The space-time parameters for the Si+Pb collisions are smaller
than those for Au+Au collisions which is consistent with the
measurements shown in Fig.~\ref{Fig:results} (a bigger source leads to a smaller
peak value of the proton correlation function). Since the number 
of participant nucleons is much larger for the Au+Au collision than
for the Si+Pb collision, a higher degree of stopping and higher baryon
 density are reached for the heavier system. The rescattering in the
 high density 
environment will not only bring the system close to the thermal
equilibrium, but also develop a strong collective velocity
field~\cite{pbm95}. The stronger velocity field in the heavier system
will 
naturally lead to a larger final freeze-out 
space-time extent. Note, we emphasize both space and time variables due to
the fact that the two-particle correlation function measures the
distance between particles at the instant the second particle is
emitted and this distance depends not only on the size of the source
but also on the duration of emission. It is instructive to compare the
 geometrical parameters of the proton 
source in the model with those of the target and projectile nuclei.
The one-dimensional  RMS charge radii of the Si and Au projectiles are
 1.6 and 3.1 fm, respectively. 
From Table~\ref{Tab:rqmd}, one can see that all transverse radii
(rms value of x-dimension in the Table) are significantly larger than
the projectile 
values for both Si+Pb and Au+Au colliding systems.  This observation is
consistent with the data from two-pion correlation measurements
~\cite{Barrette94,Barrette97} for both colliding systems, which
lends further strong support to the interpretation that transverse
expansion has occured in both reactions.  
It is also consistent with the collective transverse expansion scenario
emerging from the study of the systematics of the single-particle
transverse momentum distributions \footnote{For transverse momentum
distributions of light particles ($\pi$, K, p) it has been found 
that the slope parameter is a linear function of particle's mass,
implying a hydrodynamic-like  collective expansion and such dependence
is the strongest at
mid-rapidity.}~\cite{pbm95,na44_slope,E802qm96,E877qm96,pbm_qm97}. Note that  
the largest values of the radii in Table~\ref{Tab:rqmd} are found at
mid-rapidity, consistent with the expectation that, since the highest
particle density and collective velocity are both 
at mid-rapidity, the largest transverse expansion is expected to take
place at midrapidity.
\section{Conclusion}
 In summary, we have reported  proton correlation functions from Si+Pb
and Au+Au central collisions.  
Model calculations based on RQMD predictions coupled with Koonin-Pratt
formalism agree with the data for both systems.
Both data and model suggest that the space-time extent for the
Au+Au system is larger than that of the Si+Pb system. Combined with
the results of centrality dependence studies in Au+Au collisions this
confirms that the proton-proton correlation function is sensitive to
the number of participants involved in collision.
The analysis of model space time distributions leads to a picture
consistent with a strong collective expansion in ultrarelativistic
heavy ion collisions.  
\section{Acknowlegments}
We thank the BNL AGS and tandem operations staff, and
Dr. H. Brown for providing the various beams. We also thank
Dr. S. Pratt for providing the computer code for the calculation of the 
correlation. We are grateful to Dr. H. Sorge for making
available to us the RQMD source code.
This work was supported in part by the U.S. DoE, the NSF, NSERC,
Canada, and CNPq, Brazil.

\newpage
%
%
\begin{table}

\caption{Mean values of transverse momentum and rapidity of the proton
pairs measured for different colliding systems. See description in the
text. The units for $p_t$ are GeV/c.}   
\label{Tab:mean_val}
\begin{tabular}{ll|c|c|c} 
System & Parameter &$q_{inv}\le$0.05 GeV/c &$q_{inv}\le$0.1 GeV/c
&$q_{inv}\le$1.0 GeV/c \cr  \hline
Si+Pb                & $<p_t>$ & 0.16 & 0.18 & 0.21 \\
                     & $<y>$   & 2.18 & 2.32 & 2.34 \\ \hline
Au+Au ($16\%$-$11\%$)&$<p_t>$  & 0.35 & 0.37 & 0.33 \cr 
                     &$<y>$    & 2.56 & 2.59 & 2.51 \\ \hline
Au+Au ($4\%$)        &$<p_t>$  & 0.35 & 0.36 & 0.33 \\
                     & $<y>$   & 2.46 & 2.50  & 2.44 \cr 
\end{tabular}

\end{table}
\begin{table}
\caption{Parameters of the proton source calculated using the RQMD
model for the Si+Pb and Au+Au systems for different rapidity
intervals. All source parameters are in fm.}   
\label{Tab:rqmd}
\begin{tabular}{crcccc}
System & Source Parameters &
\multicolumn{2}{c}{$-1<y_{NN}<1$}&\multicolumn{2}{c}{In Acceptance} \\\hline 
	 &           & Mean & RMS & Mean & RMS  \\\cline{3-6}
      	 & $X$   & 0.0  & 3.9 & 0.0  & 2.5  \\
Si+Pb 	 & $Z$   & -4.2 & 6.5 & 5.1  & 5.8  \\
	 & $c\tau$   & 14.  & 7.3 & 8.4  & 6.9  \\\hline
      	 & $X$   & 0.0  & 5.7 & 0.0  & 4.7  \\
Au+Au    & $Z$   & 0.0  & 7.6 & 12.7 & 10.5 \\
         & $c\tau$   & 18.9 & 8.8 & 18.3 & 13.0 \\

\end{tabular}

\end{table}
\begin{figure}
\epsfxsize = 6.0 true in 
\epsfysize = 6.0 true in 
\centerline{\epsffile{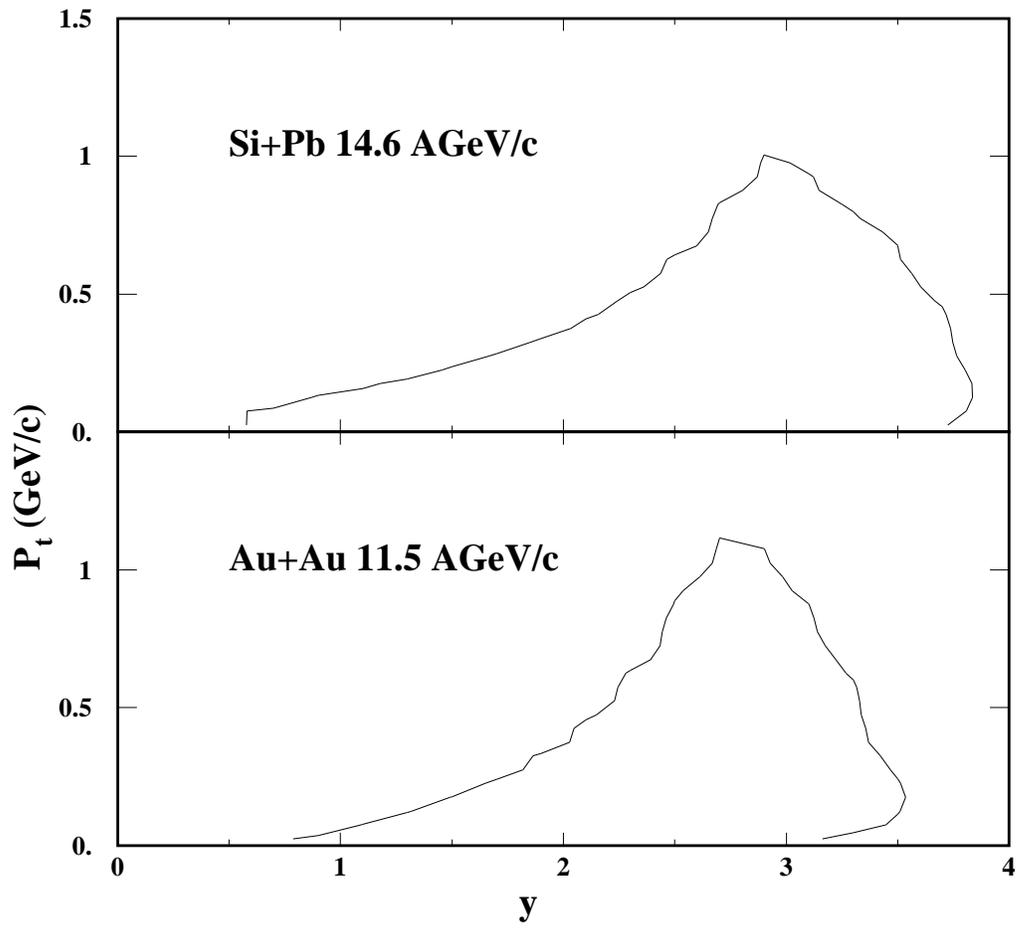}}

\caption{ E814/E877 proton acceptance in the transverse momentum -
rapidity ($p_t - y$) plane. Beam rapidities
are about 3.4 and 3.1 for Si and Au ions respectively. 
} 

\label{Fig:acceptance}
\end{figure}
\begin{figure}
\vspace{-1.0cm}
\epsfxsize = 6.0 true in 
\epsfysize = 6.0 true in 
\centerline{\epsffile{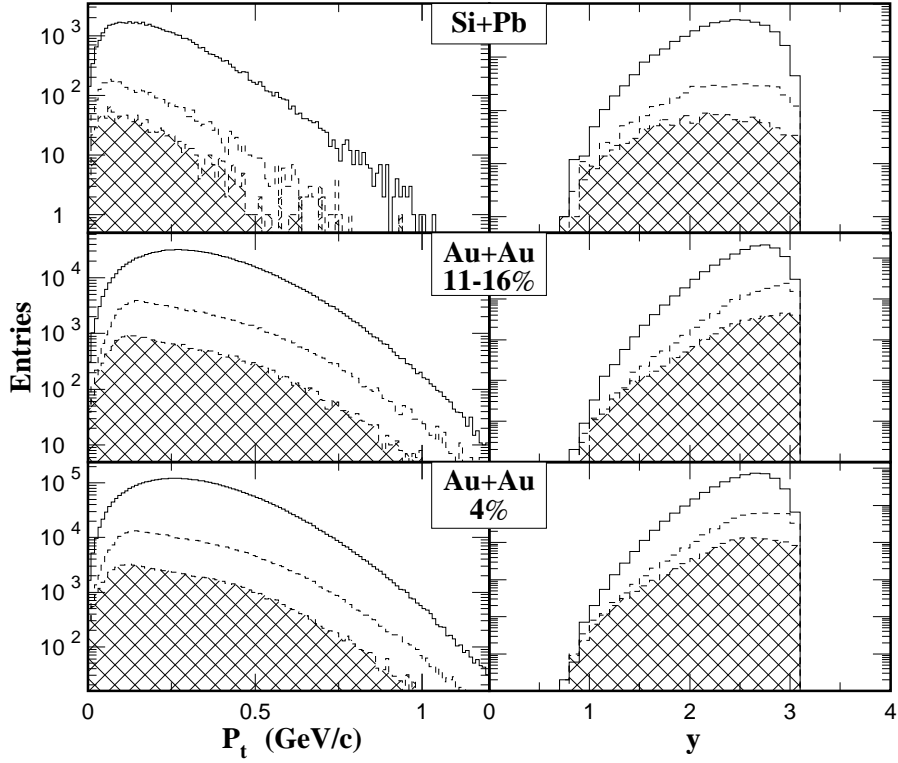}}

\caption{Transverse momentum (left panels) and rapidity (right
panels) distributions of the proton pairs for the different colliding
systems. The solid and dashed histograms represent the distributions
for pairs with $q_{inv}\le$1.0 GeV/c and $q_{inv}\le$ 0.1 GeV/c
respectively. Distributions for pairs with $q_{inv}\le$0.05 GeV/c are
hatched.}  
\label{Fig:proton_distr}
\end{figure}
\begin{figure}
\epsfxsize = 6.0 true in 
\epsfysize = 6.0 true in 
\centerline{\epsffile{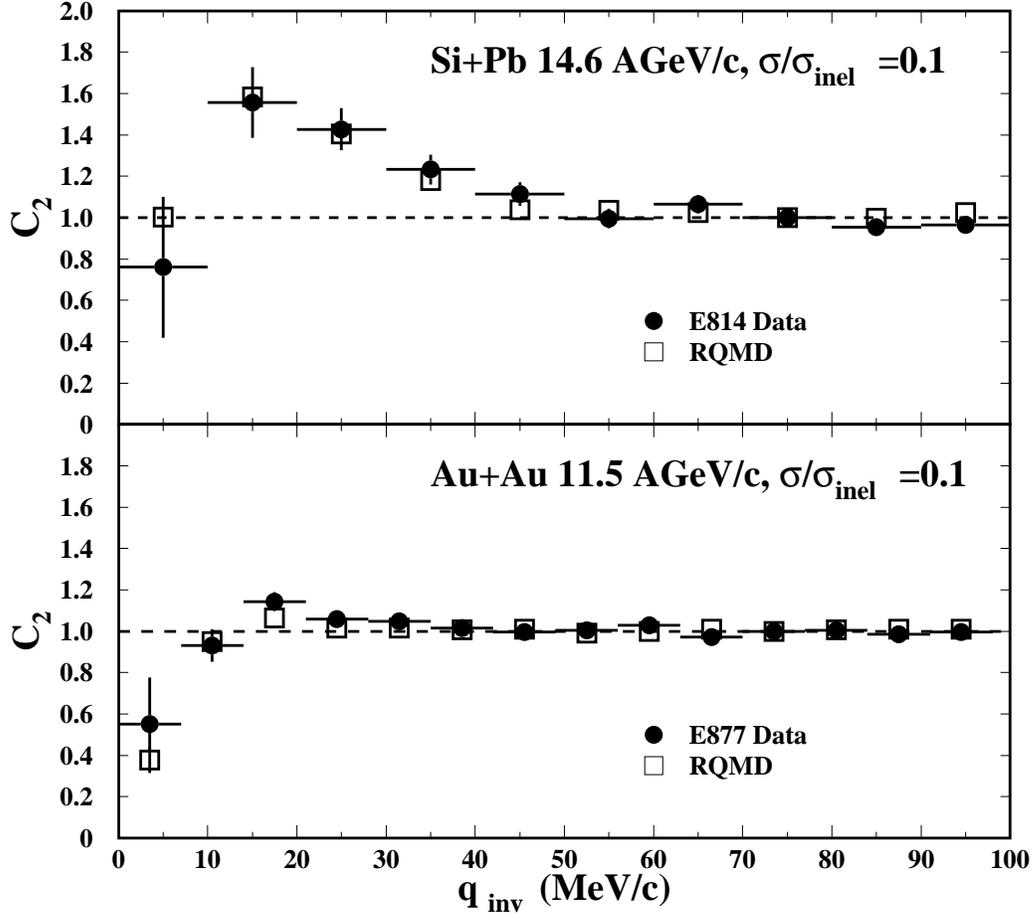}}

\caption{ Two-proton correlation functions for Si+Pb (upper panel) and
for Au+Au (lower panel) central collisions. Experimental results are
shown as filled circles while the RQMD model predictions are shown as
open squares.}
\label{Fig:results}

\end{figure}
\begin{figure}

\epsfxsize = 6.0 true in 
\epsfysize = 6.0 true in 
\centerline{\epsffile{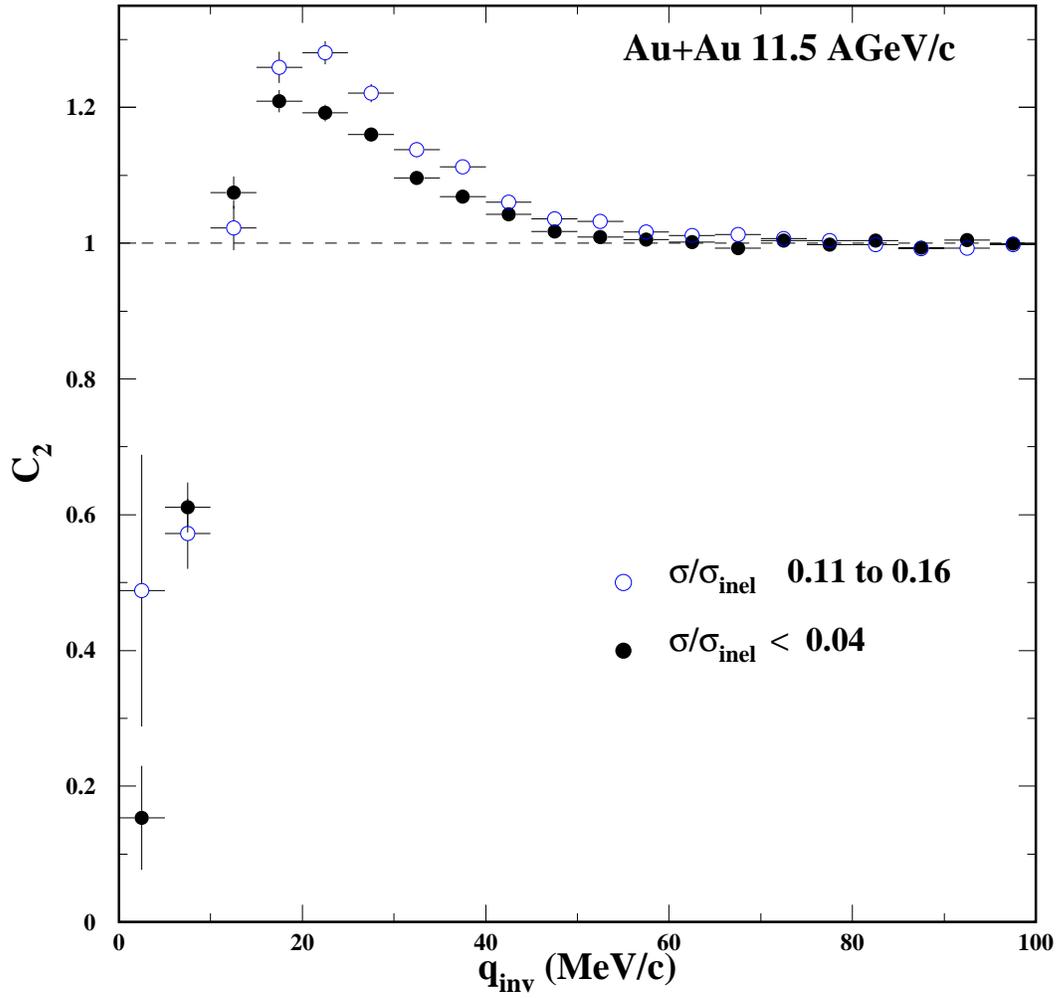}}

\caption{ Two-proton correlation functions for different centralities.} 
\label{Fig:centrality_cuts_94}
\end{figure}

\begin{figure}

\epsfxsize = 6.0 true in 
\epsfysize = 6.0 true in 
\centerline{\epsffile{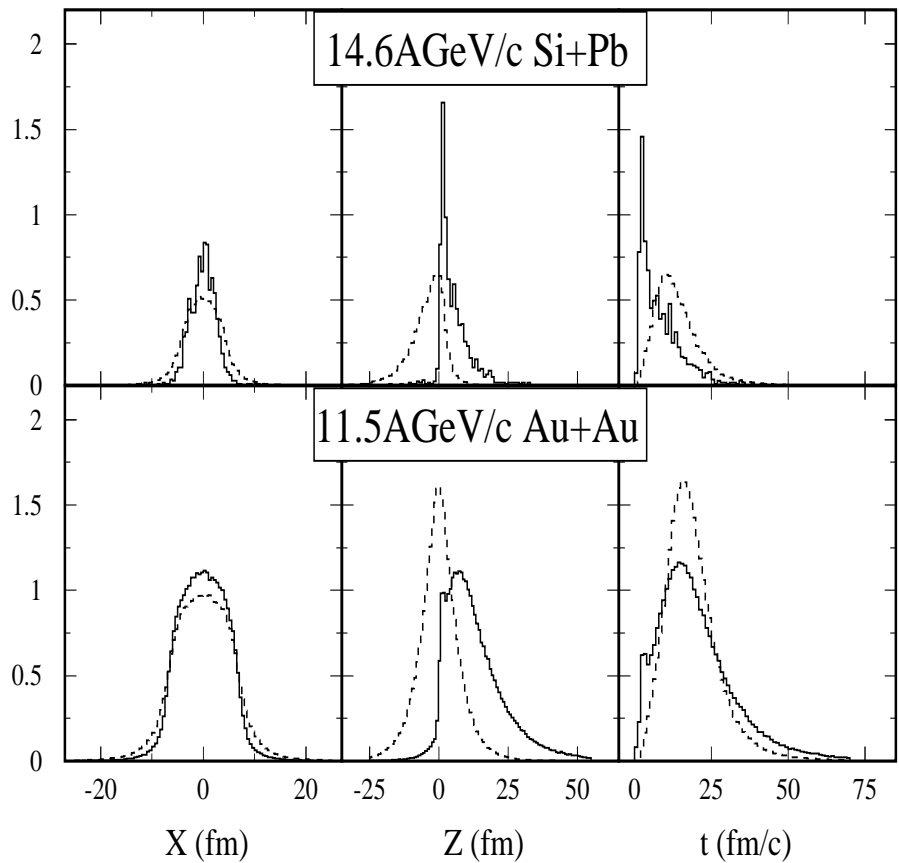}}

\caption{ Proton distributions from RQMD(v1.08). Solid lines represent
the distributions of the protons emitted into the E814/E877 spectrometer
acceptance; dashed lines represent those at mid-rapidity $-1\leq y_{NN}\leq 1$} 
\label{Fig:rqmd}

\end{figure}

\end{document}